\documentclass[11pt]{article}
\usepackage{amsfonts}
\usepackage{amsmath}
\usepackage{amssymb}
\usepackage{extarrows}
\usepackage[
  colorlinks,
  linkcolor = blue,
  citecolor = blue,
  urlcolor = blue]{hyperref}

\def \qed {\hfill \vrule height7pt width 7pt depth 0pt}

\begin{document}
 \title {On small set of one-way LOCC indistinguishability of maximally entangled states}

\author{Yan-Ling Wang$^{1}$, Mao-Sheng Li$^{2}$, Zhu-Jun Zheng$^{1}$, Shao-Ming Fei$^{3,4}$\\
 {\footnotesize  {$^1$Department of Mathematics,
 South China University of Technology, Guangzhou
510640, P.R.China}} \\
 {\footnotesize  {$^2$Department of Mathematical of Science,
  Tsinghua  University,  Beijing
100084, P.R.China}} \\
{\footnotesize{
  $^3$School of Mathematical Sciences, Capital Normal University,
Beijing 100048, China}}\\
{\footnotesize{$^4$Max-Planck-Institute for Mathematics in the Sciences, 04103
Leipzig, Germany}}
}

\maketitle

\noindent{\bf Abstract}
In this paper, we study the one-way local operations and classical communication (LOCC) problem.
In $\mathbb{C}^d\otimes\mathbb{C}^d$ with $d\geq4$, we construct a set of $3\lceil\sqrt{d}\rceil-1$
one-way LOCC indistinguishable maximally entangled states which are generalized Bell states.
Moreover, we show that there are four maximally entangled states which cannot be perfectly  distinguished by
one-way LOCC measurements for any dimension $d\geq 4$.

\noindent{\section{Introduction}}
\noindent In compound  quantum systems, many global operators  can not be implemented using only  local operations and classical communication (LOCC). This reflects  the fundamental feature of quantum mechanics called nonlocality. Meanwhile, the understanding  of the limitation of quantum operators that can be implemented by LOCC is also one of the significant subjects in quantum information theory. And local distinguishability of quantum states plays an important role in exploring quantum nonlocality \cite{Bennett99,Walgate02}. In the bipartite case, Alice and Bob share a quantum system which is chosen from one of a known set of mutually orthogonal quantum states. Their goal is to identify  the given  state using only LOCC. The nonlocality of quantum information is therefore revealed when a set of orthogonal states can not be distinguished by LOCC. Moreover, the local distinguishability has been found practical applications in quantum cryptography primitives such as secret sharing and data hiding \cite{DiVincenzo02,Markham08}.

The question of local discrimination of orthogonal quantum states has received considerable attentions in recent years
[5-19].
It is well known that any two orthogonal maximally entangled states can be perfectly distinguished with LOCC \cite{Walgate02}. In Refs.\cite{Ghosh04,Fan04},  the authors proved that a set of $d+1$ or more maximally entangled states in $d\otimes d$ systems are not perfectly locally distinguishable. Hence it is interesting to ask whether there are locally indistinguishable sets consisting of $d$ or fewer maximally entangled states in $d\otimes d$. For $d=3$, Nathanson has shown that any three maximally entangled states can be perfectly distinguished \cite{Nathanson05}. Recently, the authors in \cite{Ghosh11,Zhang} considered  one-way LOCC distinguishability and presented sets of $d$ and $d-1$ indistinguishable maximally entangled states for $d=5,...,10$. The problem remains open if there exists fewer than $d-1$ indistinguishable maximally entangled states for arbitrary dimension $d$. More recently, Nathanson showed that there exist triples of mutually orthogonal maximally entangled states in $\mathbb{C}^d\otimes\mathbb{C}^d$ which cannot be distinguished with one-way LOCC  when $d$ is even or $d \equiv 2 \ \text{ mod } 3 $ \cite{Nathanson2013}. In addition, the authors in \cite{Zhang2} gave a set with $\lceil\frac{d}{2}\rceil+2$ maximally entangled states in $\mathbb{C}^d\otimes\mathbb{C}^d$ which is  one-way LOCC indistinguishable, where $\lceil a\rceil$ means the least integer which is not less than $a$.
And in \cite{Zhang3}, the authors presented  sets with four and five maximally entangled states in $\mathbb{C}^{4m}\otimes\mathbb{C}^{4m}$  which is  one-way LOCC indistinguishable but two-way distinguishable. Whether there are four or three one-way LOCC indistinguishable maximally entangled states in arbitrary dimension remains unknown.

In this paper,  we give a positive  answer to this question  when the number of states in the set is four. First for any dimension $d\geq 4$,  we give a set of $3\lceil\sqrt{d}\rceil-1$ one-way LOCC indistinguishable maximally entangled states. Moreover, we can find four maximally entangled states  which cannot be perfectly distinguished by one-way LOCC measurements for any dimension $d\geq 4$.

\section{\bf Preliminaries}
We first introduce some basic results that will be used in proving our theorems.
Under the computational base $\{ |ij\rangle \}_{i,j=0}^{d-1}$ of  Hilbert space $\mathbb{C}^d\otimes\mathbb{C}^d $,
the generalized Bell states  are defined as follows:
\begin{equation}
|\psi_{nm}\rangle=I\otimes U_{nm}(\frac{1}{\sqrt{d}}\sum_{{j=0}}^{d-1}|jj\rangle),
\end{equation}
 where $U_{mn}=X^mZ^n$ are generalized Pauli matrices constituting a basis of unitary operators, and $X|j\rangle=|j\oplus_{d}1\rangle$, $\ Z|j\rangle=\omega^j|j\rangle$, $\omega=e^{\frac{2\pi\sqrt{ -1}}{d}}$. We define $V_{mn}=U_{mn}^T$, where $T$ stands for transpose. It is directly verified that $ZX=\omega XZ.$

 \noindent{\bf Lemma 1.} Suppose $U_{mn}=X^mZ^n, U_{m'n'}=X^{m'}Z^{n'}, $ we have $$U_{m'n'}^\dag U_{mn}=\omega^{(m'-m)n'}U_{(m-m'\text{ mod } d)(n-n'\text{ mod } d)}.$$
 \noindent\emph{Proof:}
 $$\begin{array}{ccl}
                          U_{m'n'}^\dag U_{mn}& = & (X^{m'}Z^{n'})^\dag(X^mZ^n) \\
                            & = & ({Z^\dag}^{n'}{X^\dag}^{m'})(X^mZ^n)\\
                            & = & ({Z^{{(d-1)}{n'}}}{X^{{(d-1)}{m'}}})(X^mZ^n) \\
                            & = & ({Z^{-n'}X^{-m'}})(X^mZ^n)\\
                            & = & Z^{-n'}{X^{m-m'}Z^{n}}\\
                            & = & \omega^{(m'-m)n'} X^{m-m'}{Z^{n-n'}}\\
                            & = &\omega^{(m'-m)n'}U_{(m-m'\text{mod } d)(n-n'\text{ mod } d)}.
                        \end{array}
                        $$
                        \qed

For the convenience of citation, we recall the results given in Refs.\cite{Nathanson2013,Zhang}.

 \noindent{\bf Lemma 2.} \cite{Zhang} In $\mathbb{C}^d\otimes\mathbb{C}^d $, $N\leq d$ number of pairwise orthogonal maximally entangled states $|\psi_{n_i m_i}\rangle$, $i=1,2,\ldots,N$, taken from the set given in Eq. (1), can be perfectly distinguished by one-way LOCC $A\rightarrow B$, if and only if there exists at least one state $|\alpha\rangle\in \mathcal{H}_B$ for which the states $U_{n_1 m_1}|\alpha\rangle$, $U_{n_2 m_2}|\alpha\rangle,\ldots,U_{n_N m_N}|\alpha\rangle$ are pairwise orthogonal.

 On the other hand, the set is perfectly distinguishable by one-way LOCC in the $B\rightarrow A$, if and only if there exists at least one state $|\alpha\rangle\in \mathcal{H}_A$ for which the states $V_{n_1 m_1}|\alpha\rangle,V_{n_2 m_2}|\alpha\rangle,\ldots,V_{n_N m_N}|\alpha\rangle$ are pairwise orthogonal.

\noindent{\bf Lemma 3.} \cite{Nathanson2013} Given a set of states $S=\{|\psi_i\rangle=(I\otimes U_i)|\phi\rangle \}\subset \mathbb{C}^d\otimes\mathbb{C}^d$, with $|\phi\rangle$ the standard maximally entangled state. The elements of $S$ can be perfectly distinguished with one-way LOCC if and only if there exists a set of states $\{|\phi_k\rangle\}\subset\mathbb{C}^d$ and a set of positive numbers $\{m_k\}$ such that $\sum_km_k|\phi_k\rangle\langle\phi_k|=I_d$ and $\langle\phi_k|U_j^\dag U_i|\phi_k\rangle=\delta_{ij}$ .

In the following, we concentrate ourselves on the set of maximally entangled states. Any maximally entangled state in $\mathbb{C}^d\otimes\mathbb{C}^d$
 can be written as $|\psi\rangle=(I\otimes U) |\psi_0\rangle$, where $|\psi_0\rangle=\frac{1}{\sqrt{d}}\sum_{i=1}^{d}|ii\rangle$,
and $U$  is a unitary matrix. Since there is a one to one correspondence between a maximally entangled state $|\psi_i\rangle$
and the unitary matrix $U_i$, we call the set of unitary matrices $\{U_i\}_{i=1}^{d}$ the defining unitary matrices of the set of  maximally
entangled states $\{|\psi_i\rangle\}_{i=1}^{d}$.

\section {Sets of one-way LOCC indistinguishable states}
The authors in \cite{Zhang2} presented a set with $\lceil\frac{d}{2}\rceil+2$ generalized Bell states in $\mathbb{C}^d\otimes\mathbb{C}^d$ which is  one-way LOCC indistinguishable. In the following,  firstly,  we also consider the  one-way distinguishability of generalized Bell states.

\noindent{\bf Theorem 1.}
In $\mathbb{C}^d\otimes\mathbb{C}^d~(d\geqslant4)$,  there exists an orthogonal set with $3\lceil\sqrt{d}\rceil-1$ maximally entangled states which is one-way LOCC indistinguishable:\\
$\{|\psi_{00}\rangle, |\psi_{10}\rangle, \ldots, |\psi_{n-1, 0}\rangle, |\psi_{2n-1, 0}\rangle, |\psi_{3n-1, 0}\rangle,|\psi_{4n-1, 0}\rangle, \ldots, |\psi_{(n-1)n-1, 0}\rangle, |\psi_{d-1, 0}\rangle, |\psi_{n-1, 1}\rangle,$
$|\psi_{2n-1, 1}\rangle,|\psi_{3n-1, 1}\rangle,|\psi_{4n-1, 1}\rangle,\ldots, |\psi_{(n-1)n-1, 1}\rangle, |\psi_{d-1, 1}\rangle\}$,
where $n=\lceil\sqrt{d}\rceil$.

\noindent The corresponding unitary matrices are given by\\
{
$\{U_{00}, \  \ U_{10}, \  \ \ldots, \  U_{n-1, 0},\   U_{2n-1, 0}, \ \ U_{3n-1, 0},\  U_{4n-1, 0},\  \ldots, \ \ U_{(n-1)n-1, 0},\  U_{d-1, 0},\  \ U_{n-1, 1},\  $\\
 $U_{2n-1, 1}, \  U_{3n-1, 1},\  U_{4n-1, 1},\ \ldots,\  U_{(n-1)n-1, 1},\  U_{d-1, 1}\}.$
}

\noindent\emph{Proof:}
If  $\{|\psi_{00}\rangle, |\psi_{10}\rangle, \ldots, |\psi_{n-1, 0}\rangle, |\psi_{2n-1, 0}\rangle, |\psi_{3n-1, 0}\rangle, \ldots, |\psi_{(n-1)n-1, 0}\rangle,
|\psi_{d-1, 0}\rangle, |\psi_{n-1, 1}\rangle,$\\
$|\psi_{2n-1, 1}\rangle, \ldots, |\psi_{(n-1)n-1, 1}\rangle, |\psi_{d-1, 1}\rangle\}$ can be one-way LOCC distinguished, then by lemma 2,  $\exists  \ |\alpha\rangle\neq 0 \in \mathbb{C}^{d}$,  such that the  set
 $\{U_{00}|\alpha\rangle, U_{10}|\alpha\rangle, \ldots, U_{n-1, 0}|\alpha\rangle, U_{2n-1, 0}|\alpha\rangle, U_{3n-1, 0}|\alpha\rangle,$
$  \ldots, U_{(n-1)n-1, 0}|\alpha\rangle,
U_{d-1, 0}|\alpha\rangle, U_{n-1, 1}|\alpha\rangle,
U_{2n-1, 1}|\alpha\rangle, \ldots, U_{(n-1)n-1, 1}|\alpha\rangle, U_{d-1, 1}|\alpha\rangle\} $ are  mutually orthogonal.

From the orthogonality of $U_{00}|\alpha\rangle$ and $U_{10}|\alpha\rangle, U_{20}|\alpha\rangle, \ldots, U_{n-1, 0}|\alpha\rangle,$  we obtain£º
$$\begin{array}{l}
  \langle\alpha|U_{10}|\alpha\rangle=\displaystyle\sum_{j=0}^{d-1}\omega^j\alpha_j\overline{\alpha}_j=0, \\[4mm]
  \langle\alpha|U_{20}|\alpha\rangle=\displaystyle\sum_{j=0}^{d-1}\omega^{2j}\alpha_j\overline{\alpha}_j=0, \\
  \ \ \ \ \ \  \ \ \ \ \ \ \ \ \ \ \ \ \ \vdots \\
  \langle\alpha|U_{n-1, 0}|\alpha\rangle=\displaystyle\sum_{j=0}^{d-1}\omega^{(n-1) j}\alpha_j\overline{\alpha}_j=0.
\end{array}
$$
Then by the orthogonality of $U_{2n-1, 0}|\alpha\rangle$ and $ U_{n-1, 0}|\alpha\rangle , \ldots, U_{10}|\alpha\rangle ,U_{00}|\alpha\rangle$,
taking into account with the lemma 1, we get£º
$$
\begin{array}{l}\langle\alpha|U_{n-1, 0}^{\dagger}U_{2n-1, 0}|\alpha\rangle=\langle\alpha|U_{n, 0}|\alpha\rangle=\displaystyle\sum_{j=0}^{d-1}\omega^{nj}\alpha_j\overline{\alpha}_j=0, \\
  \ \ \ \ \ \  \ \ \ \ \ \ \ \ \ \  \ \ \ \ \ \ \ \ \ \ \ \ \ \ \ \ \ \ \ \vdots \\
   \langle\alpha|U_{10}^{\dagger}U_{2n-1, 0}|\alpha\rangle=\langle\alpha|U_{2n-2, 0}|\alpha\rangle=\displaystyle\sum_{j=0}^{d-1}\omega^{(2n-2) j}\alpha_j\overline{\alpha}_j=0 ,\\[4mm]
 \langle\alpha|U_{00}^{\dagger}U_{2n-1, 0}|\alpha\rangle=\langle\alpha|U_{2n-1, 0}|\alpha\rangle=\displaystyle\sum_{j=0}^{d-1}\omega^{(2n-1) j}\alpha_j\overline{\alpha}_j=0.
\end{array}
$$

Similarly, from the orthogonality of $U_{3n-1, 0}|\alpha\rangle, U_{4n-1, 0}|\alpha\rangle, \ldots, , U_{(n-1)n-1, 0}|\alpha\rangle, U_{d-1, 0}|\alpha\rangle$ and $ U_{n-1, 0}|\alpha\rangle , \ldots, U_{10}|\alpha\rangle ,U_{00}|\alpha\rangle$,  we have:
$$\sum_{j=0}^{d-1}\omega^{(2n) j}\alpha_j\overline{\alpha}_j=\sum_{j=0}^{d-1}\omega^{(2n+1)j}\alpha_j\overline{\alpha}_j=\cdots=\sum_{j=0}^{d-1}\omega^{(d-1) j}\alpha_j\overline{\alpha}_j=0.$$
Putting the above $d-1$  equations together, we have
$$\sum_{j=0}^{d-1}\omega^{ j}\alpha_j\overline{\alpha}_j=\sum_{j=0}^{d-1}\omega^{2j}\alpha_j\overline{\alpha}_j=\sum_{j=0}^{d-1}\omega^{3j}\alpha_j\overline{\alpha}_j=\cdots=\sum_{j=0}^{d-1}\omega^{(d-1) j}\alpha_j\overline{\alpha}_j=0.$$  Solving these $d-1$  equations, we have $(\alpha_0\overline{\alpha}_0, \alpha_1\overline{\alpha}_1, \cdots, \alpha_{d-1}\overline{\alpha}_{d-1})=\lambda(1, 1, \cdots, 1).$

1) If $\lambda=0$,  then $(\alpha_0\overline{\alpha}_0, \alpha_1\overline{\alpha}_1, \cdots, \alpha_{d-1}\overline{\alpha}_{d-1})=(0, 0, \cdots, 0),$
that is,  $|\alpha\rangle=\textbf{0}.$

2) If $\lambda\neq 0$,  then for $\forall i, j, $ we have $\ \alpha_i\overline{\alpha}_j\neq 0.$
By the orthogonality of $U_{n-1, 1}|\alpha\rangle$ and $U_{n-1, 0}|\alpha\rangle, \ldots,U_{20}|\alpha\rangle, U_{10}|\alpha\rangle, U_{00}|\alpha\rangle$
and lemma 1, we have
$$\begin{array}{l}
\langle\alpha|U_{n-1, 0}^{\dagger}U_{n-1, 1}|\alpha\rangle=\langle\alpha|U_{01}|\alpha\rangle=\displaystyle\sum_{j=0}^{d-1}\omega^{0j}\alpha_j\overline{\alpha}_{j\oplus_{d}1}=0
, \\
  \ \ \ \ \ \  \ \ \ \ \ \ \ \ \ \ \ \ \ \ \ \   \  \ \ \ \ \  \ \ \ \ \ \ \vdots \\
 \langle\alpha|U_{10}^{\dagger}U_{n-1, 1}|\alpha\rangle=\langle\alpha|U_{n-2, 1}|\alpha\rangle=\displaystyle\sum_{j=0}^{d-1}\omega^{(n-2) j}\alpha_j\overline{\alpha}_{j\oplus_{d}1}=0,\\[4mm]
\langle\alpha|U_{00}^{\dagger}U_{n-1, 1}|\alpha\rangle=\langle\alpha|U_{n-1, 1}|\alpha\rangle=\displaystyle\sum_{j=0}^{d-1}\omega^{(n-1) j}\alpha_j\overline{\alpha}_{j\oplus_{d}1}=0.
\end{array}$$
By the orthogonality of $U_{2n-1, 1}|\alpha\rangle, U_{3n-1, 1}|\alpha\rangle, \ldots, U_{(n-1)n-1, 1}|\alpha\rangle, U_{d-1, 1}|\alpha\rangle$ and $U_{00}|\alpha\rangle,$\\
$ U_{10}|\alpha\rangle, U_{20}|\alpha\rangle, \ldots,
U_{n-1, 0}|\alpha\rangle$, we have
$$\sum_{j=0}^{d-1}\omega^{nj}\alpha_j\overline{\alpha}_{j\oplus_{d}1}=\sum_{j=0}^{d-1}\omega^{(n+1) j}\alpha_j\overline{\alpha}_{j\oplus_{d}1}=\cdots=\sum_{j=0}^{d-1}\omega^{(d-1) j}\alpha_j\overline{\alpha}_{j\oplus_{d}1}=0.$$
 From the above equations, $(\alpha_0\overline{\alpha}_1, \alpha_1\overline{\alpha}_2, \cdots, \alpha_{d-1}\overline{\alpha}_{0})=(0, 0, \cdots, 0)$ and $\alpha_i\overline{\alpha}_j\neq 0$ are contradictory. Therefore
$\{|\psi_{00}\rangle, |\psi_{10}\rangle, \ldots, |\psi_{n-1, 0}\rangle, |\psi_{2n-1, 0}\rangle, |\psi_{3n-1, 0}\rangle, \ldots,
|\psi_{(n-1)n-1, 0}\rangle,  $\\
$|\psi_{d-1, 0}\rangle, |\psi_{n-1, 1}\rangle, |\psi_{2n-1, 1}\rangle,\ldots, |\psi_{(n-1)n-1, 1}\rangle, |\psi_{d-1, 0}\rangle\}$ cannot be one-way LOCC distinguished.
\qed

\noindent\textbf{Remark:} It should be noticed that the above result may be worse than the known $\lceil\frac{d}{2}\rceil+2$ result \cite{Zhang2} in the case of small $d$. And $3\lceil\sqrt{d}\rceil-1\leq \lceil\frac{d}{2}\rceil+2$ when $d\geq30$, so our theorem gives a smaller one-way LOCC indistinguishable maximal entangled states in this case.

In the above discussions, we restrict ourselves on the one-way LOCC indistinguished generalized Bell states. In the following we consider
general orthogonal maximally entangled states that are indistinguishable under one-way LOCC.

\noindent{\bf Theorem 2.} There exist four  mutually orthogonal maximally entangled states in $\mathbb{C}^d\otimes\mathbb{C}^d$ which cannot be distinguished under one-way LOCC for odd  $d\geq 7$.

\emph{Proof}: Set $d=2+r$, $r\geqslant 5$. Let $P$ denote the $r\times r$ permutation matrix,
$$
P=\left[
  \begin{array}{ccccc}
    0 & 0 & \cdots & 0 & 1 \\
    1 & 0 & \cdots  & 0 & 0 \\
    0 & 1 & \cdots & 0 & 0 \\
    \vdots& \vdots & \ddots & \vdots & \vdots \\
   0& 0 & \cdots & 1 & 0 \\
  \end{array}
\right]_{r\times r}.
$$
Clearly, $P^r=I$ where $I$ denotes the $r\times r$ identity matrix.
We set $U_0=I_d$,
$$U_1=\left[
     \begin{array}{cc}
       \omega X& \\
        & P \\
     \end{array}
   \right],~~~
U_2=\left[
     \begin{array}{cc}
       \gamma Z& \\
        & P^2 \\
     \end{array}
   \right],~~~
U_3=\left[
     \begin{array}{cc}
       \sigma Y& \\
        & P^{\frac{r+1}{2}} \\
     \end{array}
   \right],
   $$
where $\omega,\gamma$ and $\sigma$ are phases satisfying $|\omega|=|\gamma|=|\sigma|=1$, $\overline{\gamma}\neq\pm i\overline{\omega}^2$,
$X,Y,Z$ are the Pauli matrices:
$$X=\left[
      \begin{array}{cc}
        0 & 1 \\
       1& 0 \\
      \end{array}
    \right],~~~
Y=\left[
      \begin{array}{cc}
        0 & -i \\
       i& 0 \\
      \end{array}
    \right],~~~
Z=\left[
      \begin{array}{cc}
        1 & 0 \\
       0& -1 \\
      \end{array}
    \right].
$$
Let $|\psi_0\rangle$ be the standard maximally entangled state, $|\psi_0\rangle=\sum_{i=0}^{d-1}|ii\rangle.$
We construct four maximally entangled states as follows:
$$
\{(I\otimes U_0)|\psi_0\rangle,(I\otimes U_1)|\psi_0\rangle,(I\otimes U_2)|\psi_0\rangle,(I\otimes U_3)|\psi_0\rangle\}\subseteq\mathbb{C}^d\otimes\mathbb{C}^d.
$$
One can check that these states are mutually orthogonal and maximally entangled.

Suppose that Alice performs an initial measurement $\mathbb{M}=\{M_k\}_{k=1}^n$ on her system and gets the measurement outcome corresponding to some operator $M_k$ $(1\leq k\leq n)$ of the following form:
$$M_k=\left[
     \begin{array}{cc}
       A_k & C_k^\dagger \\
       C_k & B_k \\
     \end{array}
   \right]\geqslant 0,$$
where $A_k$ is a $2\times 2$ matrix and $B_k$ a $r\times r$ matrix.

By lemma 3, all the measurements of Alice's can be chosen to be rank one. So we suppose all the matrices $M_k$ $(1\leq k\leq n) $ are rank one and  $M_k=|\phi_k\rangle \langle\phi_k | $ for some $|\phi_k\rangle\in \mathbb{C}^d$. In order to distinguish the above four states by one-way LOCC,  we must have $$0={\langle\phi_k |U_j^\dagger U_i|\phi_k\rangle}=Tr({U_j^\dagger U_i|\phi_k\rangle\langle\phi_k |})=Tr({U_i|\phi_k\rangle\langle\phi_k |U_j^\dagger})=Tr({U_iM_kU_j^\dagger}), \ i\neq j.$$
 That is, $Tr({U_iM_kU_j^\dagger})=0,$ whenever $ \ i\neq j$. By specify choosing $i$ and $j$, we obtain the following equations:
\begin{equation}
Tr({U_1M_k})=\omega Tr({A_kX})+Tr({B_kP})=0,
\end{equation}
\begin{equation}
Tr({U_2M_k})=\gamma Tr({A_kZ})+Tr({B_kP^2})=0,
\end{equation}
\begin{equation}
Tr({U_3M_k})=\sigma Tr({A_kY})+Tr({B_kP^{\frac{r+1}{2}}})=0,
\end{equation}
\begin{equation}
Tr({U_2M_kU_1^\dagger})=-i\overline{\omega}\gamma Tr({A_kY})+Tr({B_kP})=0,
\end{equation}
\begin{equation}
Tr({U_3M_kU_1^\dagger})=-i\overline{\omega}\sigma Tr({A_kZ})+Tr({B_kP^{\frac{r-1}{2}}})=0,
\end{equation}
\begin{equation}
Tr({U_3M_kU_2^\dagger})=-i\overline{\gamma}\sigma Tr({A_kX})+Tr({B_kP^{\frac{r-3}{2}}})=0.
\end{equation}
From equations (2) and (5), we have
\begin{equation}
\omega \,Tr({A_kX})+i\,\overline{\omega}\,\gamma\, Tr({A_kY})=0.
\end{equation}
After easily calculation, we can obtain $Tr(A_kX)=A_k(1,2)+A_k(2,1)$ and $Tr(A_kY)=i A_k(1,2)-iA_k(2,1).$ Since $A_k$ is a Hermitian matrix, then both  $Tr(A_kX)$ and $Tr(A_kY)$ are real numbers. Moving the second term of equation (8) to the right hand side then taking  the norm of each side, we have  $|Tr(A_kX)|=|Tr(A_kY)|.$ If $Tr(A_kX)\neq 0$, then we have $i\overline{\omega}^2\gamma=-\frac{Tr(A_kX)}{Tr(A_kY)}=1 \text{ or } -1.$ This is contradicted with  $\overline{\gamma}\neq \pm i\,\overline{\omega}^2.$
Hence we have $Tr(A_kX)=Tr(A_kY)=0$. Substituting $Tr(A_kY)=0$ into equation (4), we obtain $Tr({B_kP^{\frac{r+1}{2}}})=0.$
 Due to $P^r=I$ and the Hermitian of the matrix $B_k$,  the equality $Tr({B_kP^{\frac{r-1}{2}}})=\overline{Tr({B_kP^{\frac{r+1}{2}}})}$ holds, which gives rise to $Tr({B_kP^{\frac{r-1}{2}}})=0.$
Then by equation (6), we obtain $Tr({A_kZ})=0$.
Equations  $Tr(A_kX)=Tr(A_kY)=Tr(A_kZ)=0$ give that $A_k=tI_2$ for some $t_k\in \mathbb{R}$. Noticing that we have assumed $\text{rank}(M_k)=1$, so $\text{rank}(A_k)\leq 1$. Hence $A_k=\textbf{0}$ for all $1\leq k\leq n$.  But now $\sum_k^n M_{k=1}$ cannot equal to the identity $I$ for the $2\times2$ matrix of the left upper corner must equal to zero. This makes a contradiction.

Hence, we can conclude that the four states we construct above can not be distinguished by one-way LOCC.\qed

\noindent\textbf{Corollary.} There exist four  mutually orthogonal maximally entangled states in $\mathbb{C}^d\otimes\mathbb{C}^d$ which cannot be distinguished under one-way LOCC for  $d\geq 4$.

 \emph{Proof}: By the above theorem, we only need to check for the cases: $d$ is even and $d=5$. For all these cases,  it has been showed that there exist three  mutually orthogonal maximally entangled states which cannot be distinguished under one-way LOCC in Ref.\cite{Nathanson2013}. And there exists another maximally entangled state orthogonal to all the three states. So after adding such a state, these four states cannot be distinguished by one-way LOCC.
\qed

\section{Conclusion}
We study  the one-way LOCC problem and present a set of $3\lceil\sqrt{d}\rceil-1$ one-way LOCC indistinguishable maximally entangled states which are all generalized Bell states. It should be noticed that if $d$ is large enough, then the number $3\lceil\sqrt{d}\rceil-1$ is much smaller than the number $\lceil\frac{d}{2}\rceil+2$ in \cite{Zhang2}.  But for small $d$ (less than 30), our results are not so good as the known results. In addition to, we have also found four maximally entangled states  which cannot be perfectly  distinguished by one-way LOCC measurements for any dimension $d\geq 4$. For some particular dimension $d$, small one-way indistinguishable sets that contain
 only three states has been given in \cite{Nathanson2013}.  The question whether there exist three one-way indistinguishable maximally entangled states  for arbitrary $d\geq4$ remains open.
\bigskip

\noindent\textbf{Acknowledgments} We are thankful for the referees' suggestions and careful reading of our paper. This work is supported by NSFC (Grant Nos. 11475178, 11275131).

\end{document}